EUDET-Report-2008 -07

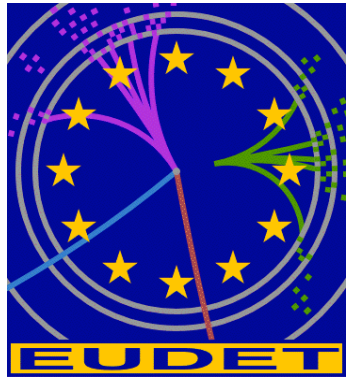

# Calibration prototype for the EUDET HCAL

J. Cvach[1], M. Janata[1], J. Kvasnička[1,2], D. Lednický[1], I. Polák[1], J. Smolík[1], J. Zálešák[1]

December 16, 2008

**Abstract**

The electronics producing a few nanoseconds long LED pulses tuneable in the amplitude and the optical system distributing the LED light to scintillators by a single optical fibre are described. The properties of the LED pulses are controlled by a computer connected via a CANbus to the controller on the board. We produced a prototype with 6 LED pulse drivers placed on a 4-layer PCB with dimension 250 x 147 mm$^2$. The pulse width is constant and 3 ns long. Its amplitude can be tuned from zero to 1.2 A. The pulse frequency can be set in the range up to 100 kHz. The LED light is distributed by a 1 mm diameter optical fibre. On the length of 2 m the fibre has every 30 mm a notch which flashes light with homogeneity of ±20 %. The system is foreseen for calibration of silicon photomultipliers embedded in scintillation tiles.

---

[1] Institute of Physics of the ASCR, v.v.i., Prague, Czech Republic
[2] Czech Technical University in Prague, Dept. of Computer Science and Engineering, Prague, Czech Republic

EUDET-Report-2008-07

# 1 Introduction

In the existing physics prototype [1] of the tile hadron calorimeter (HCAL) equipped with Silicon Photomultipliers (SiPM) [2] as photodetectors, the calibration and monitoring system distributes UV LED[3] light to each tile via one clear fibre. The system performs three different tasks:

- provides a calibration of the SiPM gain at low light intensity
- monitors all SiPMs during test beam operations with a fixed-intensity light pulse
- measures the full SiPM response function by varying the light intensity from zero to the saturation level. By varying the voltage, the LED intensity can cover the entire dynamic range of all SiPMs.

The design of the previous LED driver [3] was optimized for the generation of nearly rectangular pulses, i.e. pulses with a fast rise time and fast drop-off time. The stability of light was monitored by PIN photodiodes. Since PIN photodiodes have a gain of one, an additional charge-sensitive preamplifier for the PIN photodiode readout is needed. The presence of a high-gain preamplifier directly on the board in the vicinity of the strong signals for LEDs has turned out to be a source of crosstalk.

In the new concept of the LED driver we decided to abandon the rectangular pulses and rather use a sinusoidal pulse shape. We call it a quasi-resonant driver and it is described in Section 2. Section 3 deals with the control of the LED driver by a computer via a microcontroller. We investigated simplification of the light distribution of the physics prototype and propose a new option with notched fibres where each fibre supplies a row of tiles with calibration flashes. Section 4 deals with the light distribution system.

The calibration system described in this report is a part of wider activities aiming at the construction of a compact technological prototype [4] of the analogue HCAL. A design of another calibration system [5] using miniature LEDs sitting above each tile is developed in parallel. To diminish the electrical interference with the readout electronics this system uses only fixed amplitude for the LED to allow gain calibration at the level of several photoelectrons.

# 2 The Quasi Resonant LED Driver

Using the experience with the Calibration and Monitoring Board (CMB) and SiPMs at the HCAL we found a value of the fixed pulse-width of 5 ns as an optimum. The new system employs a LED driver with a low level of noise and parasitic high frequency spectrum. It is called the Quasi-Resonant LED Driver - QRLED. The principal circuit diagram is on the Figure 1.

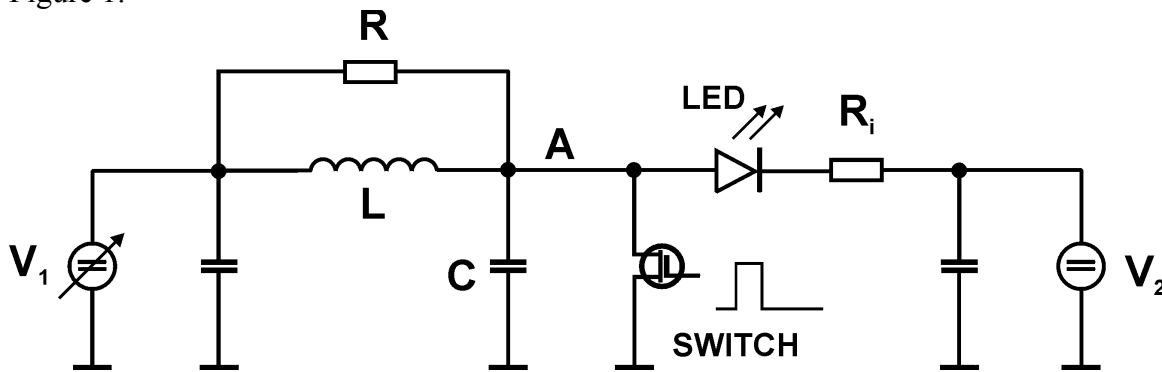

**Figure 1 The principal circuit diagram of the QRLED driver**

---
[3] in the following we use LED instead of UV LED throughout this report

The whole circuit is connected between two DC voltages, $V_2$ is constant, $V_1$ is variable but always lower than $V_2$. The main components generating the electric pulse are inductor L, capacitor C and a switch (in our case MOSFET). When the switch is on, current begins to flow through the inductor until the switch is off. The energy stored in the inductor is transferred into the capacitor C and the voltage across the capacitor (point A) is rising up very fast. The slope of the rise is defined by the value of the serial resonant circuit LC and is around 2 ns (for L = 50 nH, C = 22 pF). When all charge from the inductor is transferred, the rise of the voltage stops and the discharging of the capacitor begins. If the LED is connected to the point A, the discharging current flows through it and the LED starts to shine and the voltage at the point A falls down (see Figure 2). Since the cathode of the LED is connected via the resistor $R_i$ to the constant voltage $V_2$, the LED shines only if the voltage in the point A is higher than $V_2$. By setting the voltage $V_1$ it is possible to change the amplitude of the output light pulse. The current flowing through the LED is seen as a voltage drop across the resistor $R_i$, 1 Ω in our case (see Figure 1). At the end of the electric pulse an oscillation starts. It uses the remaining charge on the capacitor after the LED is reversely polarized and the current flow is blocked. These oscillations can create more than one output pulses in each period. To suppress this unwanted effect, a dumping resistor R is connected parallel to the inductor. With the optimal value of the resistor it is possible to keep the oscillations below the opening voltage of the LED and the optical LED output pulse is only one in each period.

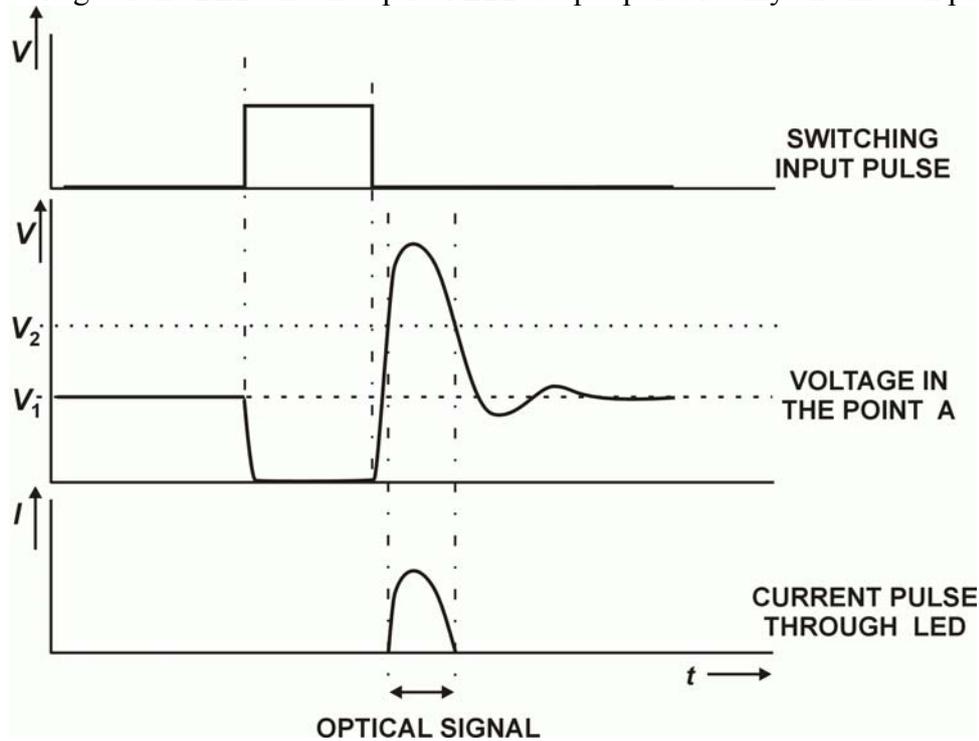

**Figure 2  Correspondence between electrical and optical pulses in the QRLED driver**

The intensity of the optical pulse is proportional to the stored energy in the inductor. On the multichannel board, we have to trim voltage $V_2$, common to all QRLED drivers, to make an amplitude scale. The voltage $V_1$ is tuned individually to control the stored energy in an inductor which is proportional to the amplitude of the LED pulse.
On the 4-layer PCB we developed an integrated toroidal inductor with non magnetic core. This inductor has about 50 nH inductance. There are individual turns of the inductor made by Cu traces on the top and the bottom layers of the PCB, connected by a standard vias (see Figure 3). This inductor provides relevant temperature stability. We tested to freeze parts on the PCB by a freezing spray (to ~ -20° C) and observed a shift in position of the pulse by 1 ns at maximum. An amplitude change is around of a few percents.

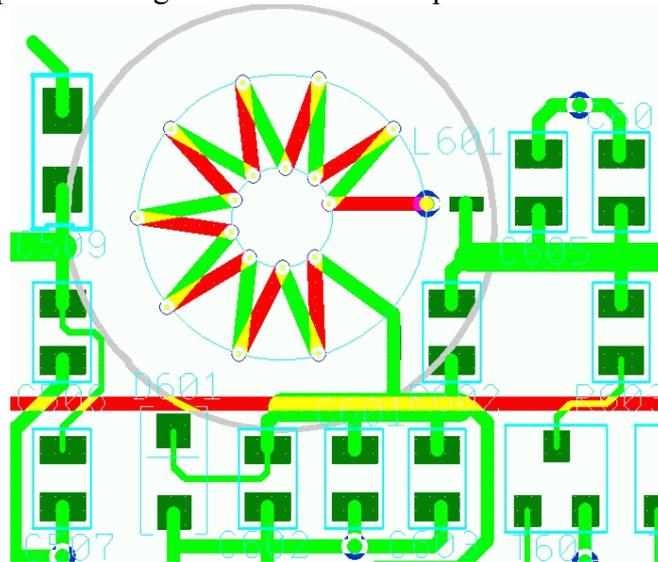

**Figure 3  Detail of the inductor design on the PCB**

## 3   Control of the Driver

The board is controlled by an AT91SAM7X256 microprocessor which provides a human interface bridge between the CANbus (or DIP-switches) and the control of six QRLED drivers. The architecture of the board is shown in Figure 4. Each QRLED driver is driven by a trigger signal, the LED voltage $V_1$ and the common LED bias voltage $V_2$. All variable voltages are generated by a 12-bit DAC (AD5328) with eight channels. Six DAC channels drive amplifier boosting voltages $V_1$ and one DAC channel generates common bias voltage $V_2$ for all QRLED drivers. The last DAC channel is connected back to ADC for testing and calibration. The DAC is controlled over the SPI bus.

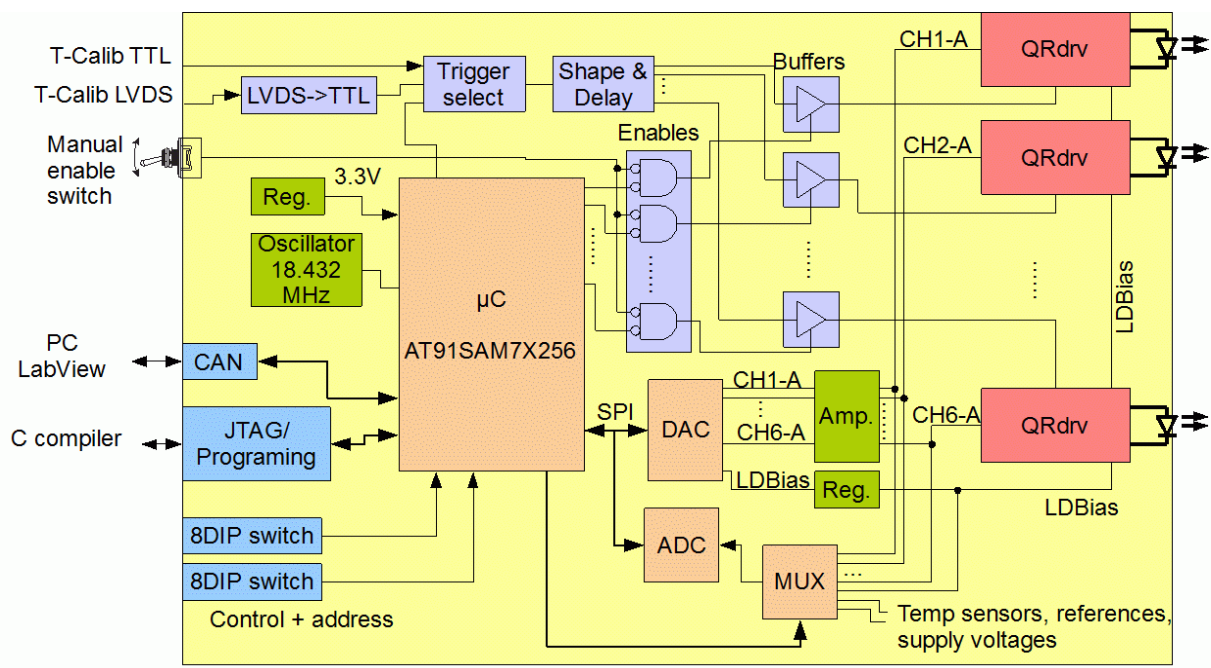

**Figure 4  The principal scheme of the QRLED driver**

The board provides measurements of 16 different voltage signals. All signals are multiplexed by analogue multiplexer (ADG706) to the sigma-delta ADC (AD7792). The range of each input is adapted to the 16 ADC range by a voltage dividers. The ADC is controlled over the SPI bus and the multiplexer is controlled by the direct processor signal.

One internal and two external trigger sources are available. The internal trigger is generated by the processor. The external trigger (T-Calib) signals are derived from the LVDS and TTL levels, respectively. The trigger source is selected by onboard jumper. The trigger signal is reshaped and delayed for each QRLED channel separately. The trigger signal to each channel is driven by hi-current logic buffers and gated by one global enable signal and by a per-channel enable signal. A common enable signal is generated by the manual toggle switch. Each channel has a separate enable signal driven by the processor. The LED flashes only when both enable signals are active.

## 3.1 Modes of operation

The board can be operated in 2 modes: stand-alone (without PC) and CANbus modes. Operation mode is chosen by one of 16 DIP switches, the other fifteen have different meaning in each mode.

### 3.1.1 Stand-alone mode

In the stand-alone mode, a limited control is provided by 15 DIP switches. The channels are enabled by 6 DIP-switches. The bias voltage $V_2$ can be configured by 2 DIP-switches to 4 levels. The control voltage $V_1$ of all 6 channels is set at once by 5 DIP-switches to 32 levels. The internal trigger can be configured by 2 DIP-switches to 4 frequencies.

### 3.1.2 CANbus mode

The CANbus mode provides a full control over the trigger frequency, DAC and provides ADC readout. The communication is performed by CAN 2.0A Standard Frames at 125 kbits/s. The board address is set by the DIP-switches and multiplied by two by the system. On this address the board accepts packets. The board transmits packets at the address incremented by one.

Communication protocol is based on sending commands and requests from PC to the QRLED driver board. Each successful execution of a CAN command is followed by a corresponding response from the board. After a setting of a subset of measured signals, the ADC measurements are sent asynchronously after each successful conversion.

## 3.2 Firmware

The processor is programmed (or updated) by a 20-pin JTAG connector. The system is based on the FreeRTOS real-time operating system. Depending on the operating mode different processes are running. A simple switch-read process runs in stand-alone mode. In the CANbus mode, all messages are read by the interrupt routine and passed to the main process. The ADC readout occupies an independent process. Incoming and outgoing messages are queued. Firmware is coded in C and compiled by ARM port of GCC.

## 3.3 Control Application

The control application of the QRLED board is written in LabView and utilizes the Kvaser Canlib library. The application provides
- user interface to QRLED driver board (shown at Figure 5)
- enable and disable of each channels separately

- setting of all voltages
- selection of a subset for measurement
- display of measured results
- graph of a history of temperature variation
- indication of a loss of communication (caused for example by the reset or a power cycle).

The configuration of the board can be resynchronized in both directions by writing the actual GUI configuration to the board or by acquiring the board configuration into the GUI.

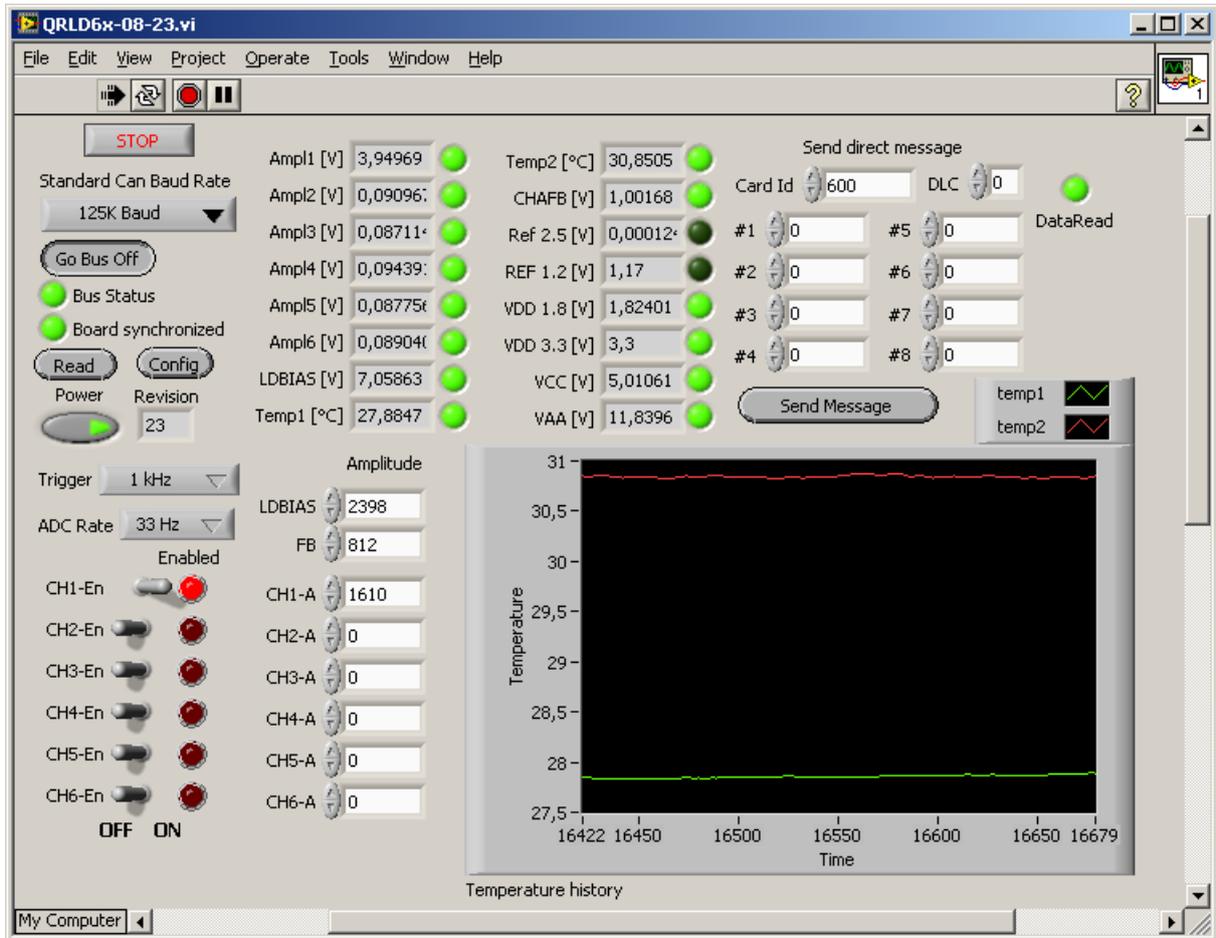

**Figure 5** LabView panel of the control of the QRLED board

## 4  Light Distribution System

For the application in a more complex HCAL prototype we developed a new system for distribution of calibration light. In the physics prototype each tile had its own clear fibre. Eighteen fibres were coupled to one LED. For thousands tiles it leads to a complicated system with significant space requirements. Therefore we started to think about a new system with smaller space requirements. We decided to replace the concept "one fibre - one tile" by the "one fibre - many tiles" solution. To distribute calibration light by one fibre to many tiles we need a fibre which radiates light in side direction. This can be achieved by a controlled damage of a standard fibre (we call it a "notched fibre") or using special commercial "side-emitting" fibres.

To use commercial fibres seems to be the easiest solution. We tested three types of fibres with a diameter of 1 mm from two producers. The measured dependence of amplitude of the side emitted light on distance is shown in Figure 6. As we can see, there is a strong (approximately exponential) decreasing dependence on the distance. The ratio of the emitted light to the light passing through the fibre is very low. Due to the two reasons these fibres are not suitable for our needs.

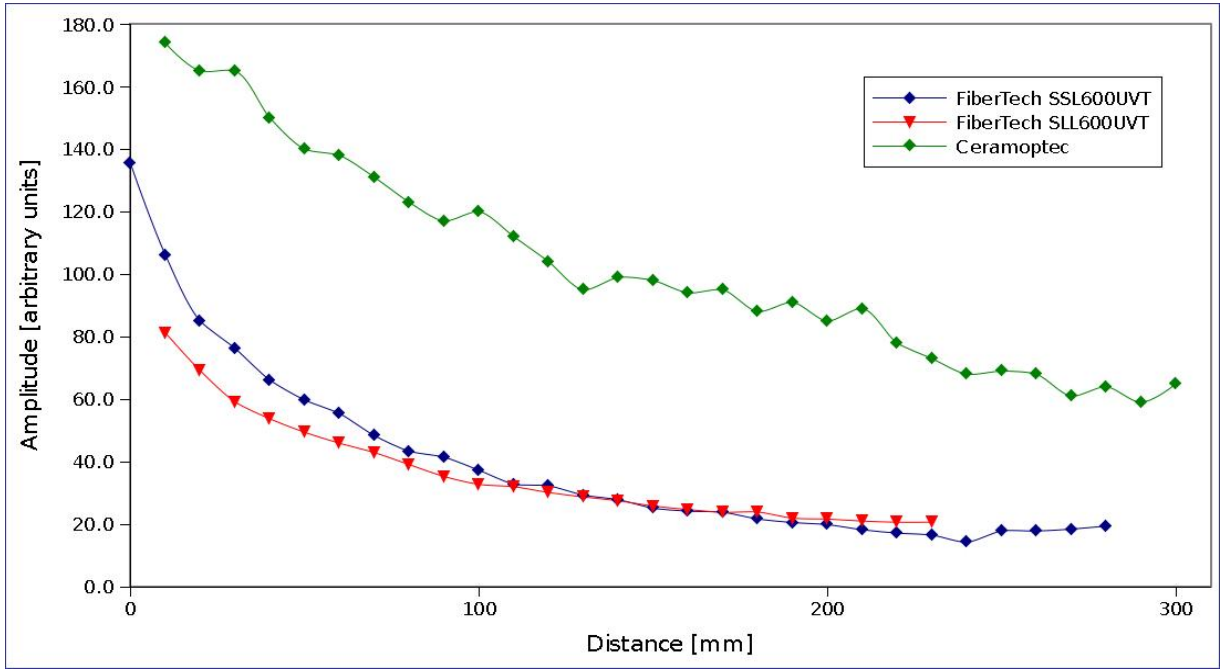

Figure 6  Measured dependence of light emitted from commercial side-emitting fibres on distance

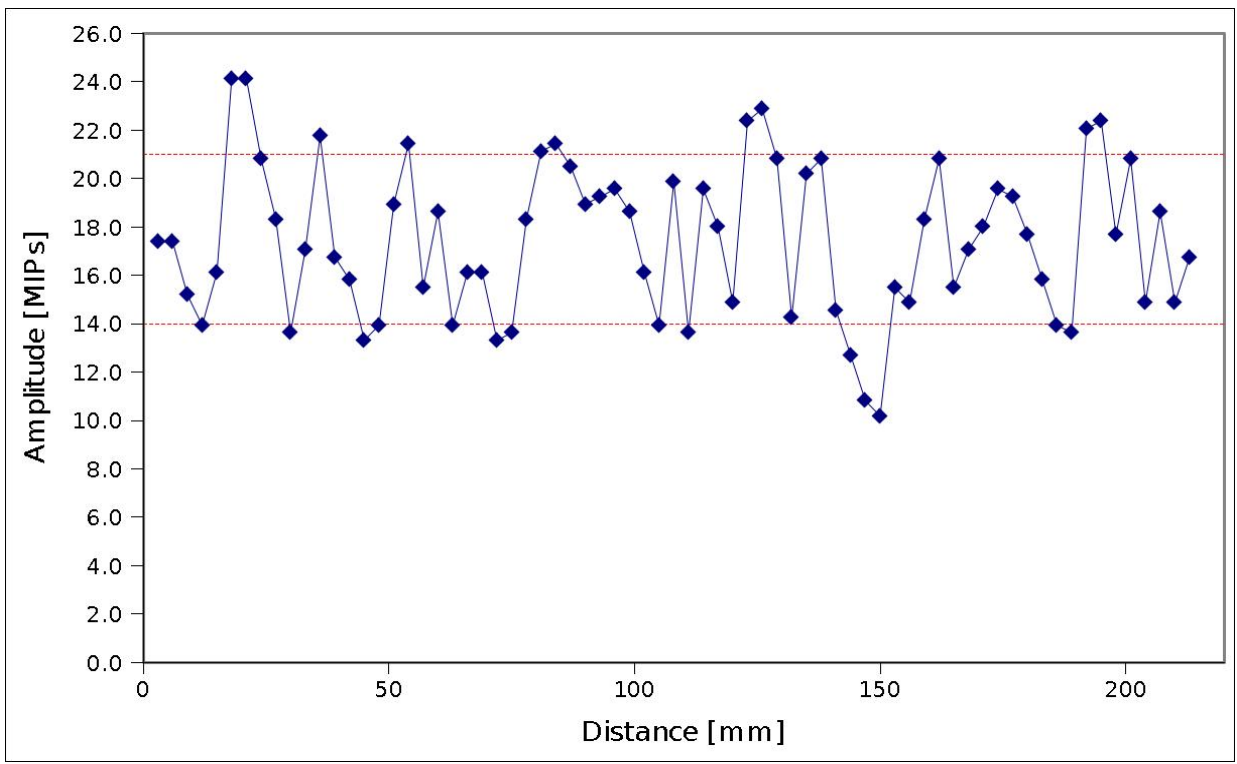

Figure 7  Measured dependence of light emitted from the first prototype of the "notched" fibre on distance. The dashed lines correspond to the 15 % dispersion

Thus we decided to develop fibres with notches. The starting requirements were a fibre with 70 shining points in the distance of 3 cm with weak dependence of the emitted light on the distance and small variance (±15%) in the amplitude of the emitted light. The first fibre prototype did not meet all our requirements but was very close to them (Figure 7). It was demonstrated that the fibre with a good uniformity is possible to produce. Now the second prototype is in production and it seems to fulfil our initial requirements. The prototypes are hand-made and for the mass production a machined process must be developed. We can expect that it will bring further improvement in uniformity of light distribution. The R&D of the "notched" fibres is done in the collaboration with the Safibra company [6].

The solution where many tiles are connected by one fibre can cause light cross-talk. The scintillation light produced in one tile can spread by the fibre into the adjacent tiles. This effect can decrease the performance of the HCAL. Therefore we measured the crosstalk and found that it is invisible in our test stand.

The first prototype showed that the maximal signal generated in tiles approaches 20 MIPs. The simulation showed that the maximal occupancy in a tile corresponds to the level of about 100 MIPs. This can be hardly achieved by a modification of fibre. The possible ways are:

- extended width of the calibration pulse
- use of the LED with higher light power and/or smaller radiating angle
- improvement of the coupling between LED and fibre.

We have tested several types of LEDs with wavelength around 400 nm. The results are shown in Tables 1 for LEDs with the diameter of 5 mm and Table 2 with the diameter of 3 mm. The wavelength was determined using minispectrometer Hamamatsu TM UV/VIS C10082CA-01 with spectral response range from 200nm to 800nm. Signal amplitudes were calculated as average value of signals from three tiles with SiPM. The light pulses were generated with the QRLED driver prototype and distributed by a notched fibre into tiles. The viewing angle corresponds to the vertex angle of the visible light cone emitted from LED and detected on the screen. We can see that some LED candidates can bring an improvement at level of tens percents compared to the default LED (the one used in the calibration system of the physics prototype [1]). We observe a fast development in the LED segment of market, thus a further improvement in LED output power can be expected soon.

A very important part of the optical system is coupling between LED and fibre. The fibre with a diameter of 1 mm retains only small part of light radiated from LED with diameter of 5 mm. Therefore, we are studying the possibilities of application of an optical coupler.

We shall continue in the development of the system with "notched" fibre as we think that it is feasible and it can reach sufficient performance.

**Table 1 Tested 5 mm LEDs**

| Producer specification | | | | | | Our measurements | | |
|---|---|---|---|---|---|---|---|---|
| Catalogue Number (*) | Wavelength λ (nm) | Output power | Viewing angle (deg.) | Vf (V) | If (A) | Wavelength peak λ (nm) | Signal amplitude (A.U) | Viewing angle (deg) |
| default | - | - | - | - | - | 405; 805 | 2400 | 22 |
| 4150-127 | blue | 8 cd | 20 | 3.2 | 20 | 470 | 1150 | 32 |
| 4150-158 | white | 18 cd | 15 | 3.5 | 20 | 450; 540 | 1250 | 28 |
| 511-878 CreeChip | 390-395 | 12 mW | 30 | 3.6 | 20 | 410; 810 | 2350 | 46 |
| LED 375-6 | 375 | 2.5 mW | 6 | 3.0 | 30 | 370 | 1950 | 6 |

**Table 2 Tested 3 mm LEDs**

| Producer specification | | | | | | Our measurements | | |
|---|---|---|---|---|---|---|---|---|
| Catalogue Number (*) | Wavelength $\lambda$ (nm) | Output power | Viewing angle (deg.) | Vf (V) | If (A) | Wavelength peak $\lambda$ (nm) | Signal amplitude (A.U.) | Viewing angle (deg) |
| 4150-140 | white | 1.7 cd | 15 | 3.0 | 20 | 465 | 2000 | 19 |
| 4150-171 | 400 | 1.0 cd | 30 | 3.5 | 20 | 410 | 3000 | 33 |
| 511-773 CreeChip | 390-395 | 12 mW | 30 | 3.5 | 20 | 400 | 3300 | 36 |

*\* LEDs with catalogue number beginning with 4150 can be found at www.tipa.eu, with 511 at www.gme.cz and LED 375-6 is from www.roithner-laser.com.*

# 5 Performance

We produced a functional sample of the QRLED driver and the first tests were done. The results correspond to our expectations and requirements. This prototype of 6 channels QRLED calibrator is built on the four-layer PCB (see Figure 8). All components are placed on the same side of the PCB. The QRLED drivers have the same shape and they are placed along the longer edge of the PCB. Other circuits, as a microprocessor control, trigger generator in range of 1 Hz to 100 kHz, the power regulators and the PIN photodiode preamplifiers are located in the centre and on the far side. We placed on the board the HCAL T-calib interface, which can be used as external trigger.

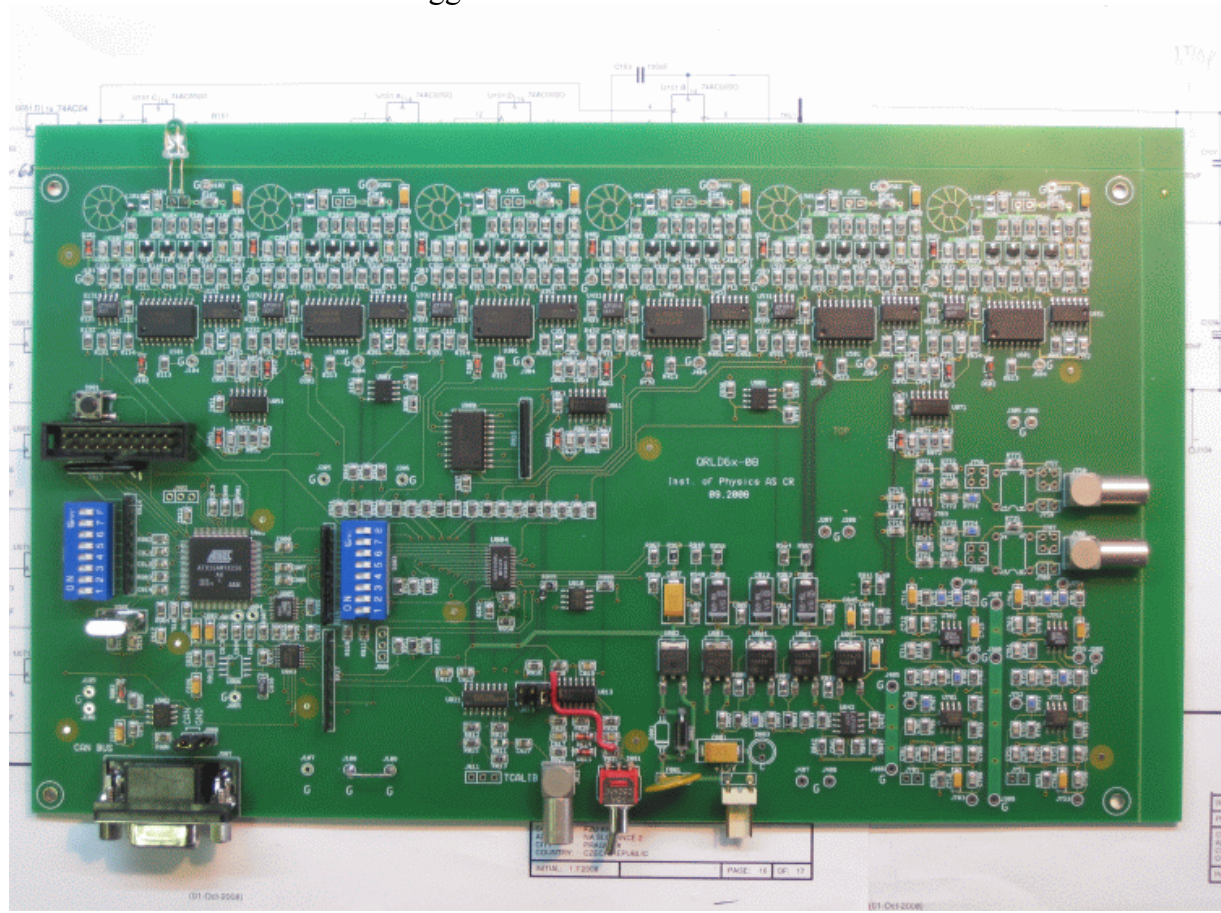

**Figure 8 View of the six channel QRLED placed on the 4-layer PCB board**

The original goal was to generate pulse 6 ns wide. It appeared that the printed toroidal inductors had lower inductance than expected. As a result the final current pulse-width is 2.5 ns only. This value was measured with help of a differential 1 GHz probe and 1 GHz oscilloscope TDS4104. To reach our goal of 6 ns, the inductor has to have larger inductance around 120 nH.

In the oscillograms shown in Figure 9 the cyan trace represents the amplified signal from the PIN diode, the green trace gives the voltage at point A (Figure 1). The yellow trace corresponds to the current flowing through resistor $R_i$. The oscillograms show small and large current, which is primer to the output amplitude. They were measured with a voltage 1 GHz differential probe on the 1 Ω smd 0805 resistor. The forward LED current is positive over the baseline, marked 1 at the left side. Note the horizontal scale is 10 ns/div and the vertical is 500 mV/div ( corresponds to the current of 500mA ).

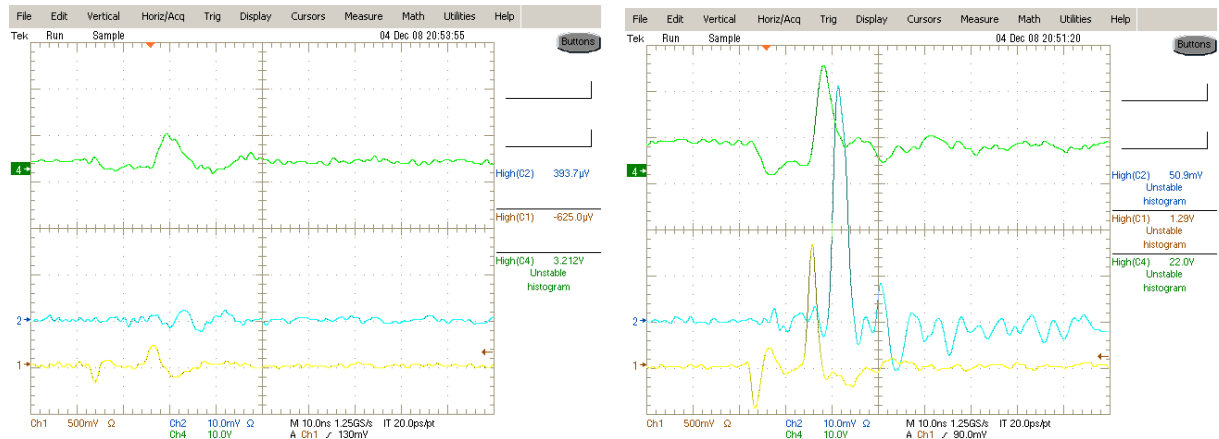

**Figure 9  Oscillograms of the electrical pulses generating LED light by the QRLED driver. Left oscillogram is for small amplitude, the right one for the maximal amplitude of the pulse**

The LED has a certain parallel parasitic capacitance. A current passing through the LED is split between this capacitance and the active electrical to optical conversion process in chip. This can be seen at the oscillograms. Conforming test will be done with proved fast photodetector (PM, SiPM).

In next prototype, we shall make optimization of the circuits with regard to the miniaturization and increase the inductance of the inductor to get longer LED pulses.

# 6  Conclusion

We constructed a prototype of six channel UV LED driver which meets basic requirements for calibration of SiPM photodetectors:
- sinusoidal pulse shape with several nanoseconds pulse width
- tuneable pulse amplitude
- drivers are controlled from external PC via LabView.

The prototype is complemented by the optical system that distributes light from the UV LED by a single "notched" fibre to a row of scintillator tiles. The uniformity of the light radiated by notches is constant within ±20 %.

The future improvement on the electronic part will concentrate on further miniaturization of circuits and increase of the UV LED pulse width to 5 – 6 ns. The improvements in the optical system we see in the increase of the light output from fibre notches by a factor up to five by better optical coupling LED – fibre, more powerful UV LEDs and better homogeneity of the emitted light from notches.

## Acknowledgement

This work is supported by the Commission of the European Communities under the 6$^{th}$ Framework Programme "Structuring the European Research Area", contract number RII3-026126 and by the Ministry of Education of the Czech Republic under the projects LC527 and INGO-1P05LA259.